\pdfoutput=1
\documentclass[12pt,a4paper]{article}

\usepackage{ifthen} 
\newboolean{pdflatex}
\setboolean{pdflatex}{true} 

\newboolean{articletitles}
\setboolean{articletitles}{true} 

\newboolean{uprightparticles}
\setboolean{uprightparticles}{false} 

\def\paperauthors{LHCb collaboration} 
\def\paperasciititle{Branching fraction measurement of the decay B+ -> psi(2S) phi(1020) K+} 
\def\papertitle{Branching fraction measurement of the decay $B^+ \to \psi(2S) \phi(1020) K^+$} 
\def\paperkeywords{{High Energy Physics}, {LHCb}} 
\def\papercopyright{\the\year\ CERN for the benefit of the LHCb collaboration} 
\def\paperlicence{CC BY 4.0 licence}
\def\paperlicenceurl{https://creativecommons.org/licenses/by/4.0/}

\usepackage{xspace} 
\usepackage{upgreek}

\def\DeltaOrDeltabar  {\kern 0.18em\optbar{\kern -0.18em \PDelta}{}\xspace}
\def\XiOrXibar        {\kern 0.18em\optbar{\kern -0.18em \PXi}{}\xspace}
\def\SigmaOrSigmabar  {\kern 0.18em\optbar{\kern -0.18em \PSigma}{}\xspace}
\def\OmegaOrOmegabar  {\kern 0.18em\optbar{\kern -0.18em \POmega}{}\xspace}

\def\GeVcc {\gevcc}

\newcommand{\TeVnosp}{\ifthenelse{\boolean{inbibliography}}{\ensuremath{~T\kern -0.05em eV}}{\ensuremath{\mathrm{Te\kern -0.1em V}}}}
\newcommand{\GeVnosp}{\ensuremath{\mathrm{Ge\kern -0.1em V}}}
\newcommand{\MeVnosp}{\ensuremath{\mathrm{Me\kern -0.1em V}}}
\newcommand{\keVnosp}{\ensuremath{\mathrm{ke\kern -0.1em V}}}
\newcommand{\eVnosp}{\ensuremath{\mathrm{e\kern -0.1em V}}}
\newcommand{\GeVcnosp}{\ensuremath{{\mathrm{Ge\kern -0.1em V\!/}c}}}
\newcommand{\MeVcnosp}{\ensuremath{{\mathrm{Me\kern -0.1em V\!/}c}}}
\newcommand{\GeVccnosp}{\ensuremath{{\mathrm{Ge\kern -0.1em V\!/}c^2}}}
\newcommand{\MeVccnosp}{\ensuremath{{\mathrm{Me\kern -0.1em V\!/}c^2}}}
\newcommand{\GeVGeVccccnosp}{\ensuremath{{\mathrm{Ge\kern -0.1em V^2\!/}c^4}}}

\def\BtoJpsiphiKDecay{{\ensuremath{B^{+}\to J/\psi\phi K^{+}}}\xspace}
\def\Jpsitomumu{{\ensuremath{J/\psi\to \mu^{+}\mu^{-}}}\xspace}
\def\psitomumu{{\ensuremath{\psi(2S)\to \mu^{+}\mu^{-}}}\xspace}
\def\BtopsiphiKDecay{{\ensuremath{B^{+}\to\psi(2S)\phi K^{+}}}\xspace}

\def\phitokk{{\ensuremath{\phi \to K^+ K^-}\xspace}}

\usepackage[top=1in, bottom=1.25in, left=1in, right=1in]{geometry}

\usepackage{microtype}
\usepackage{lineno}  
\usepackage{xspace} 
\usepackage{caption} 

\usepackage{graphicx}  
\usepackage{color}
\usepackage{colortbl}
\graphicspath{{./figs/}} 

\usepackage{amsmath} 
\usepackage{amssymb}
\usepackage{amsfonts}
\usepackage{upgreek} 

\newcommand*\patchAmsMathEnvironmentForLineno[1]{%
\expandafter\let\csname old#1\expandafter\endcsname\csname #1\endcsname
\expandafter\let\csname oldend#1\expandafter\endcsname\csname
end#1\endcsname
 \renewenvironment{#1}%
   {\linenomath\csname old#1\endcsname}%
   {\csname oldend#1\endcsname\endlinenomath}%
}
\newcommand*\patchBothAmsMathEnvironmentsForLineno[1]{%
  \patchAmsMathEnvironmentForLineno{#1}%
  \patchAmsMathEnvironmentForLineno{#1*}%
}
\AtBeginDocument{%
\patchBothAmsMathEnvironmentsForLineno{equation}%
\patchBothAmsMathEnvironmentsForLineno{align}%
\patchBothAmsMathEnvironmentsForLineno{flalign}%
\patchBothAmsMathEnvironmentsForLineno{alignat}%
\patchBothAmsMathEnvironmentsForLineno{gather}%
\patchBothAmsMathEnvironmentsForLineno{multline}%
\patchBothAmsMathEnvironmentsForLineno{eqnarray}%
}

\usepackage{hyperxmp}

\usepackage[pdftex,
            pdfauthor={\paperauthors},
            pdftitle={\paperasciititle},
            pdfkeywords={\paperkeywords},
            pdfcopyright={Copyright (C) \papercopyright},
            pdflicenseurl={\paperlicenceurl}]{hyperref}

\usepackage[colorinlistoftodos,textsize=scriptsize]{todonotes}

\usepackage[bottom,flushmargin,hang,multiple]{footmisc}

\usepackage[all]{hypcap} 

\usepackage{xspace} 
\usepackage{upgreek}

\def\lhcb   {\mbox{LHCb}\xspace}

\def\MagUp {\mbox{\em Mag\kern -0.05em Up}\xspace}

\ifthenelse{\boolean{uprightparticles}}%
{

 \def\Ppsi        {\ensuremath{\uppsi}\xspace}

 \def\PDelta      {\ensuremath{\Delta}\xspace}                 
 \def\PXi         {\ensuremath{\Xi}\xspace}                 
 \def\PLambda     {\ensuremath{\Lambda}\xspace}                 
 \def\PSigma      {\ensuremath{\Sigma}\xspace}                 
 \def\POmega      {\ensuremath{\Omega}\xspace}                 
 \def\PUpsilon    {\ensuremath{\Upsilon}\xspace}

 \def\PB      {\ensuremath{\mathrm{B}}\xspace}                 
                  
 \def\PD      {\ensuremath{\mathrm{D}}\xspace}

 \def\PJ      {\ensuremath{\mathrm{J}}\xspace}                 
 \def\PK      {\ensuremath{\mathrm{K}}\xspace}

 \def\Pb      {\ensuremath{\mathrm{b}}\xspace}                 
 \def\Pc      {\ensuremath{\mathrm{c}}\xspace}

 \def\Pi      {\ensuremath{\mathrm{i}}\xspace}

 \def\Ps      {\ensuremath{\mathrm{s}}\xspace}

 \def\thebaroffset{0.0em}
}
{

 \def\Ppsi        {\ensuremath{\psi}\xspace}                 
                  
 \mathchardef\PDelta="7101
 \mathchardef\PXi="7104
 \mathchardef\PLambda="7103
 \mathchardef\PSigma="7106
 \mathchardef\POmega="710A
 \mathchardef\PUpsilon="7107
                  
 \def\PB      {\ensuremath{B}\xspace}                 
                  
 \def\PD      {\ensuremath{D}\xspace}

 \def\PJ      {\ensuremath{J}\xspace}                 
 \def\PK      {\ensuremath{K}\xspace}

 \def\Pb      {\ensuremath{b}\xspace}                 
 \def\Pc      {\ensuremath{c}\xspace}

 \def\Pi      {\ensuremath{i}\xspace}

 \def\Ps      {\ensuremath{s}\xspace}

 \def\thebaroffset{0.18em}
}
\newcommand{\offsetoverline}[2][\thebaroffset]{\kern #1\overline{\kern -#1 #2}}%

\makeatletter
\ifcase \@ptsize \relax
  \newcommand{\miniscule}{\@setfontsize\miniscule{4}{5}}
\or
  \newcommand{\miniscule}{\@setfontsize\miniscule{5}{6}}
\or
  \newcommand{\miniscule}{\@setfontsize\miniscule{5}{6}}
\fi
\makeatother

\DeclareRobustCommand{\optbar}[1]{\shortstack{{\miniscule (\rule[.5ex]{1.25em}{.18mm})}
  \\ [-.7ex] $#1$}}

\def\squark    {{\ensuremath{\Ps}}\xspace}

\def\cquark    {{\ensuremath{\Pc}}\xspace}

\def\bquark    {{\ensuremath{\Pb}}\xspace}

\def\kaon    {{\ensuremath{\PK}}\xspace}

\def\KorKbar {\kern \thebaroffset\optbar{\kern -\thebaroffset \PK}{}\xspace}

\def\Kp      {{\ensuremath{\kaon^+}}\xspace}
\def\Km      {{\ensuremath{\kaon^-}}\xspace}

\def\D       {{\ensuremath{\PD}}\xspace}

\def\DorDbar {\kern \thebaroffset\optbar{\kern -\thebaroffset \PD}\xspace}

\def\Dp      {{\ensuremath{\D^+}}\xspace}
\def\Dm      {{\ensuremath{\D^-}}\xspace}

\def\DpDm    {\ensuremath{\Dp {\kern -0.16em \Dm}}\xspace}

\def\B       {{\ensuremath{\PB}}\xspace}

\def\BorBbar {\kern \thebaroffset\optbar{\kern -\thebaroffset \PB}\xspace}

\def\Bd      {{\ensuremath{\B^0}}\xspace}

\def\BdorBdbar {\kern \thebaroffset\optbar{\kern -\thebaroffset \Bd}\xspace}
\def\Bu      {{\ensuremath{\B^+}}\xspace}

\def\Bp      {{\ensuremath{\Bu}}\xspace}

\def\Bs      {{\ensuremath{\B^0_\squark}}\xspace}

\def\BsorBsbar {\kern \thebaroffset\optbar{\kern -\thebaroffset \Bs}\xspace}

\def\jpsi     {{\ensuremath{{\PJ\mskip -3mu/\mskip -2mu\Ppsi}}}\xspace}

\def\Y#1S{\ensuremath{\PUpsilon{(#1S)}}\xspace}

\def\LorLbar     {\kern \thebaroffset\optbar{\kern -\thebaroffset \PLambda}\xspace}

\def\BF         {{\ensuremath{\mathcal{B}}}\xspace}
\def\BR         {\BF}

\def\to                 {\ensuremath{\rightarrow}\xspace}

\def\AT#1     {\ensuremath{A_{\mathrm{T}}^{#1}}\xspace}           

\def\C#1      {\ensuremath{\mathcal{C}_{#1}}\xspace}                       
\def\Cp#1     {\ensuremath{\mathcal{C}_{#1}^{'}}\xspace}                    
\def\Ceff#1   {\ensuremath{\mathcal{C}_{#1}^{\mathrm{(eff)}}}\xspace}        
\def\Cpeff#1  {\ensuremath{\mathcal{C}_{#1}^{'\mathrm{(eff)}}}\xspace}       
\def\Ope#1    {\ensuremath{\mathcal{O}_{#1}}\xspace}                       
\def\Opep#1   {\ensuremath{\mathcal{O}_{#1}^{'}}\xspace}                    

\newcommand{\nospaceunit}[1]{\ensuremath{\text{#1}}}       
\newcommand{\aunit}[1]{\ensuremath{\text{\,#1}}}       

\newcommand{\tev}{\aunit{Te\kern -0.1em V}\xspace}
\newcommand{\gev}{\aunit{Ge\kern -0.1em V}\xspace}
\newcommand{\mev}{\aunit{Me\kern -0.1em V}\xspace}
\newcommand{\kev}{\aunit{ke\kern -0.1em V}\xspace}
\newcommand{\ev}{\aunit{e\kern -0.1em V}\xspace}
 
\newcommand{\mevc}{\ensuremath{\aunit{Me\kern -0.1em V\!/}c}\xspace}
\newcommand{\gevc}{\ensuremath{\aunit{Ge\kern -0.1em V\!/}c}\xspace}
\newcommand{\mevcc}{\ensuremath{\aunit{Me\kern -0.1em V\!/}c^2}\xspace}
\newcommand{\gevcc}{\ensuremath{\aunit{Ge\kern -0.1em V\!/}c^2}\xspace}

\def\mum  {\ensuremath{\,\upmu\nospaceunit{m}}\xspace}

\def\fb   {\ensuremath{\aunit{fb}}\xspace}
\def\invfb   {\ensuremath{\fb^{-1}}\xspace}

\newcommand{\chisq}{\ensuremath{\chi^2}\xspace}
\newcommand{\chisqndf}{\ensuremath{\chi^2/\mathrm{ndf}}\xspace}
\newcommand{\chisqip}{\ensuremath{\chi^2_{\text{IP}}}\xspace}

\def\gsim{{~\raise.15em\hbox{$>$}\kern-.85em
          \lower.35em\hbox{$\sim$}~}\xspace}
\def\lsim{{~\raise.15em\hbox{$<$}\kern-.85em
          \lower.35em\hbox{$\sim$}~}\xspace}

\def\sPlot{\mbox{\em sPlot}\xspace}

\def\pt         {\ensuremath{p_{\mathrm{T}}}\xspace}

\def\ptot       {\ensuremath{p}\xspace}

\def\evtgen     {\mbox{\textsc{EvtGen}}\xspace}

\def\geant      {\mbox{\textsc{Geant4}}\xspace}

\def\photos     {\mbox{\textsc{Photos}}\xspace}

\def\pythia     {\mbox{\textsc{Pythia}}\xspace}

\def\tell1  {TELL1\xspace}
\def\ukl1   {UKL1\xspace}

\def\sPlot{\mbox{\em sPlot}\xspace}

\newcommand{\lhcborcid}[1]{\href{https://orcid.org/#1}{\hspace*{0.1em}\raisebox{-0.45ex}{\includegraphics[width=1em]{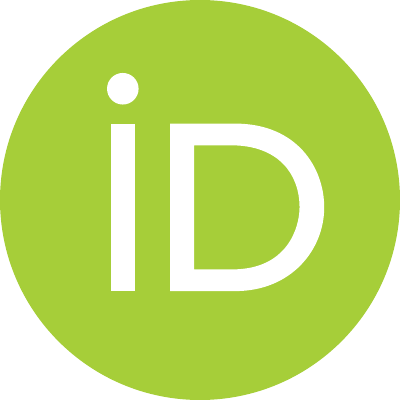}}}}
\usepackage{cite} 
\usepackage{mciteplus}

\usepackage{longtable} 
\usepackage{multirow}
\usepackage{color}

\definecolor{brickred}{rgb}{0.8, 0.25, 0.33}

\begin{document}

\renewcommand{\thefootnote}{\fnsymbol{footnote}}
\setcounter{footnote}{1}


\begin{titlepage}
\pagenumbering{roman}

\vspace*{-1.5cm}
\centerline{\large EUROPEAN ORGANIZATION FOR NUCLEAR RESEARCH (CERN)}
\vspace*{1.5cm}
\noindent
\begin{tabular*}{\linewidth}{lc@{\extracolsep{\fill}}r@{\extracolsep{0pt}}}
\ifthenelse{\boolean{pdflatex}}
{\vspace*{-1.5cm}\mbox{\!\!\!\includegraphics[width=.14\textwidth]{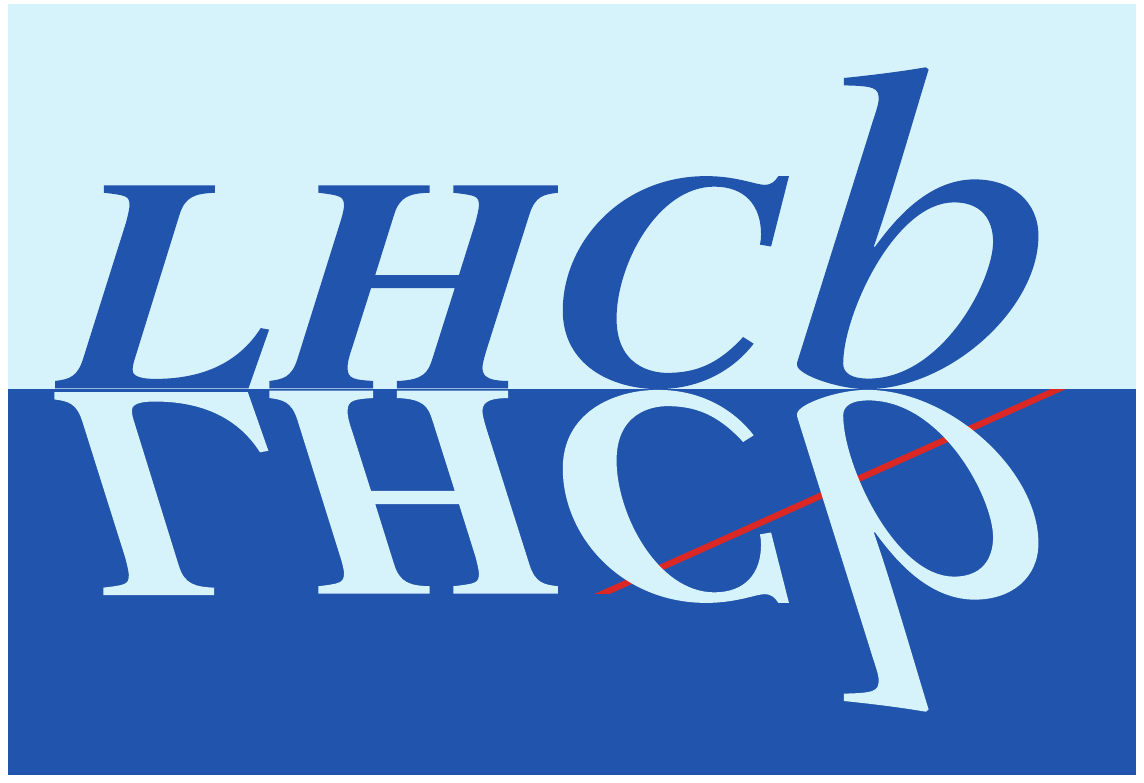}} & &}%
{\vspace*{-1.2cm}\mbox{\!\!\!\includegraphics[width=.12\textwidth]{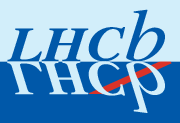}} & &}%
\\
 & & CERN-EP-2025-011 \\  
 & & LHCb-PAPER-2024-039 \\  
 & & May 14, 2025 \\ 
 & & \\
\end{tabular*}

\vspace*{2.0cm}

{\normalfont\bfseries\boldmath\huge
\begin{center}
  \papertitle 
\end{center}
}

\vspace*{2.0cm}

\begin{center}
\paperauthors\footnote{Authors are listed at the end of this paper.}
\end{center}

\vspace{\fill}

\begin{abstract}
  \noindent
   The branching fraction of the decay
   $B^+\to \psi(2S)\phi(1020)K^+$, 
   relative to the topologically similar decay 
   $B^+\to J/\psi \phi(1020) K^+$,
   is measured using proton-proton collision data
   collected by the LHCb experiment at center-of-mass energies of 7, 8, and 13 TeV, corresponding to an integrated luminosity of $9\,\mathrm{fb}^{-1}$. 
   The ratio is found to be $0.061 \pm 0.004 \pm 0.009$, where the first uncertainty is statistical and the second systematic.
   Using the world-average branching fraction
   for $B^+ \to J/\psi \phi(1020) K^+$, the 
   branching fraction for the decay  
   $B^+\to \psi(2S) \phi(1020) K^+$
   is found to be $ (3.0 \pm 0.2 \pm 0.5 \pm 0.2) \times 10^{-6} $, 
   where the first uncertainty is statistical, the second systematic, and the third is due to the branching fraction  of the normalization channel.  

\end{abstract}

\vspace*{2.0cm}

\begin{center}
  Published in Phys.~Rev.~D 111 (2025) 092008 
\end{center}

\vspace{\fill}

{\footnotesize 
\centerline{\copyright~\papercopyright. \href{\paperlicenceurl}{\paperlicence}.}}
\vspace*{2mm}

\end{titlepage}


\newpage
\setcounter{page}{2}
\mbox{~}


\renewcommand{\thefootnote}{\arabic{footnote}}
\setcounter{footnote}{0}

\cleardoublepage


\pagestyle{plain} 
\setcounter{page}{1}
\pagenumbering{arabic}


\section{Introduction}
\label{sec:Introduction}
The 
$B^+ \to \psi (2S) \phi K^+$ 
decay
proceeds through the same quark transition 
$b\to c\bar{c}s$
as the topologically similar $B^+ \to \jpsi \phi K^+$ decay.\footnote{Throughout this paper inclusion of the charge-conjugated decays is implied and the $\phi(1020)$ is referred to as $\phi$.}  In the latter, states compatible with exotic  
mesons have
been observed, such as the 
$\chi_{c1}(4140)$ state~\cite{CDF:2009jgo,CMS:2013jru,D0:2013jvp,LHCb-PAPER-2016-019} 
and the $T_{c \bar{c} \bar{s} 1}(4000)$ state~\cite{LHCb-PAPER-2020-044}.\footnote{In this paper the new PDG naming conventions~\cite{PDG2024} for exotic hadrons are used. 
The states formerly known as the 
$X(4140)$,  
$Z_{cs}^+(4000)$ and
$X(4700)$
are now referred to as the 
$\chi_{c1}(4140)$, $T_{c \bar c \bar s 1}(4000)$ and $\chi_c(4700)$, respectively.}
The $B^+ \to \psi(2S) \phi K^+ $ phase space is about 80 times smaller than that of the 
 $ B^+ \to \jpsi \phi K^+  $ decay, while the branching fraction
${\cal B}(B^+ \to \psi(2S) \phi K^+ )$ is
 about 15 times smaller than that of the 
$B^+ \to \jpsi \phi K^+$ decay~\cite{Khachatryan:2016npv}.
The average amplitude of the former process is 
therefore greater than
that of the latter process.
Thus, the $ B^+ \to \psi (2S) \phi K^+ $ decay provides the potential to make observations of high-mass resonant contributions that are complementary to those observed in the decay $ B^+ \to \jpsi \phi K^+$,
such as the $\chi_c(4700)$ state~\cite{Dias:2013qga,Voloshin:2019ilw}.    

This paper presents a 
measurement of the branching fraction of the decay 
\mbox{$B^+ \to\psi(2S) \phi K^+$} 
relative to the normalization decay $ B^+ \to \jpsi \phi  K^+ $.
The intermediate resonances are reconstructed as \psitomumu, \Jpsitomumu, and \phitokk.
This analysis uses the full Run~1 (2011--2012) and Run~2 (2015--2016, 2017, and 2018) LHCb data samples of proton-proton ($pp$) collisions at center-of-mass energies of 7, 8, and 13~TeV, corresponding to an 
integrated luminosity of about $9 \invfb$.
The branching fraction of \BtopsiphiKDecay, calculated using the $\BR(\BtoJpsiphiKDecay) \nolinebreak$ value provided by the PDG~\cite{PDG2024}, is also reported. 
This LHCb result is compared with a measurement of the branching fraction of $B^+ \to \psi (2S) \phi K^+ $ 
reported by the CMS collaboration~\cite{Khachatryan:2016npv}.

\section{Detector and simulation}
\label{sec:Detector}
 
The \lhcb detector~\cite{LHCb-DP-2008-001,LHCb-DP-2014-002} is a single-arm forward
spectrometer covering the \mbox{pseudorapidity} range $2<\eta <5$,
designed for the study of particles containing \bquark or \cquark
quarks. The detector includes a high-precision tracking system
consisting of a silicon-strip vertex detector surrounding the $pp$
interaction region~\cite{LHCb-DP-2014-001}, a large-area silicon-strip detector located
upstream of a dipole magnet with a bending power of about
$4{\mathrm{\,T\,m}}$, and three stations of silicon-strip detectors and straw
drift tubes~\cite{LHCb-DP-2013-003,LHCb-DP-2017-001}
placed downstream of the magnet.
The tracking system provides a measurement of the momentum, \ptot, of charged particles with
a relative uncertainty that varies from 0.5\% at low momentum to 1.0\% at 200\gevc.
The minimum distance of a track to a primary vertex (PV), the impact parameter (IP), 
is measured with a resolution of $(15+29/\pt)\mum$,
where \pt is the component of the momentum transverse to the beam, in\,\gevc.
Different types of charged hadrons are distinguished using information
from two ring-imaging Cherenkov detectors~\cite{LHCb-DP-2012-003}. 
Photons, electrons, and hadrons are identified by a calorimeter system consisting of
scintillating-pad and preshower detectors, an electromagnetic
and a hadronic calorimeter. Muons are identified by a
system composed of alternating layers of iron and multi-wire
proportional chambers~\cite{LHCb-DP-2012-002}.
The online event selection is performed by a trigger~\cite{LHCb-DP-2012-004}, 
which consists of a hardware stage, based on information from the calorimeter and muon
systems, followed by a two-level software stage, which applies a full event reconstruction.
Between the two software stages, an alignment and calibration of the detector is performed in near real-time and their results are used in the trigger~\cite{LHCb-PROC-2015-011}.
The same alignment and calibration information is propagated 
to the offline reconstruction software to ensure consistency
and avoid additional systematic uncertainties that would
be required were they different.

Simulation is required to model the effects of the detector acceptance and the
selection requirements.
In the simulation, $pp$ collisions are generated using
\pythia~\cite{Sjostrand:2007gs,*Sjostrand:2006za} 
with a specific \lhcb configuration~\cite{LHCb-PROC-2010-056}.
Decays of unstable particles
are described by \evtgen~\cite{Lange:2001uf}, in which final-state
radiation is generated using \photos~\cite{davidson2015photos}.
The interaction of the generated particles with the detector, and its response,
are implemented using the \geant
toolkit~\cite{Allison:2006ve, *Agostinelli:2002hh} as described in
Ref.~\cite{LHCb-PROC-2011-006}. 

\section{Branching fraction measurement}
\label{sec:BR}

The ratio of branching fractions ${\cal R}_{\rm BF}$ is measured as 
\begin{equation}
{\cal R}_{\rm BF} \equiv
  \frac{\BR(\BtopsiphiKDecay)}{\BR(\BtoJpsiphiKDecay)} = \frac{N_{\rm{Signal}}}{N_{\rm{Norm}}} 
  \frac{F_{{\rm{Signal}}}}{F_{{\rm{Norm}}}} 
  \frac{\epsilon_{\rm{Norm}}}{\epsilon_{\rm{Signal}}} \frac{\BR(\Jpsitomumu)}{\BR(\psitomumu)} \, ,
\label{eq:BF}
\end{equation}
where $\BR(\Jpsitomumu)=(5.961 \pm 0.033)\%$ and $\BR(\BtoJpsiphiKDecay)=(5.0 \pm 0.4) \times 10^{-5}$ are the world-average branching fractions~\cite{PDG2024}.
Assuming lepton flavor universality of the electroweak interaction, 
the branching fraction of the $\psitomumu$ decay can be replaced by that of the $\psi(2S)\to e^+e^-$ decay, equal to $(0.793 \pm 0.017)\%$~\cite{PDG2024}, which has an uncertainty approximately three times smaller.  
The number of $B^+$ decays for the signal (normalization) mode
with a charmonium decay combined with three kaons, including sidebands for the  $\phi $ candidates in the  $K^+K^-$ mass spectra, 
is denoted by $N_{\rm{Signal(Norm)}}$. This is measured by fitting the $B^+$ meson mass distribution. 
The fraction of candidates in the $K^+K^-$ invariant mass spectrum attributed to the $\phi$ decays
is denoted by 
$F_{{\mathrm{Signal(Norm)}}}$, 
and is obtained by fitting the $K^+K^-$  mass distribution.
Finally, $\epsilon_{\rm{Signal(Norm)}}$ is the efficiency of the signal (normalization) channel, and is obtained from simulation.  

\subsection{Selection}
\label{sec:Selection}

Candidate \BtopsiphiKDecay  
signal and \BtoJpsiphiKDecay normalization decays are selected using a hardware trigger, 
a two-stage software trigger, and offline event reconstruction criteria followed by two  Boosted Decision Tree (BDT)~\cite{Breiman,AdaBoost} classifiers that
suppress background significantly.
Selection criteria are identical for both channels, with the exception of the dimuon mass windows for the 
$\psi$, where $ \psi $
denotes either the
$ \psi(2S) $ or the $\jpsi$ meson.

{The hardware trigger selects events with at least one high-\pt muon candidate later associated with the 
$ \psi $ in the \Bp decay or selects
events independently of any track used in the decay.}
The first stage of the software trigger selects a pair of oppositely charged muons with high 
invariant mass or a muon with high \pt and high \chisqip 
for each track (or composite particle). 
The variable \chisqip is defined as 
the difference in the vertex-fit \chisq of the primary-vertex fit reconstructed with and without the track under consideration. 
The second stage of the software trigger selects pairs of muons with invariant mass within 120\mevcc of the $\psi$ mass 
that form vertices which are
significantly detached from the closest primary vertex.

Offline, \Bp meson candidates are formed from combinations
of $ \psi $ candidates with three kaons where one of the $ \Kp\Km $ pairs has an invariant mass between 1000\mevcc and 1040\mevcc.
The $ \psi $ windows are tightened to $\pm 100\mevcc$ around the nominal masses
which induces negligible losses of efficiency.
Kaon candidates are required to have \pt larger than 100\mevc, 
$\chisqip > 1$, 
and to have particle-identification signatures compatible with the kaon hypothesis.
The  $\phi$ candidates are formed from pairs of oppositely charged kaons that originate from a common vertex and have \pt greater than 200\mevc.
The  \Bp candidates are required to have good vertices, to have invariant masses within 60\mevcc of the nominal \Bp mass, and to have momentum
vectors
aligned with the \Bp line of flight.
A vertex fit is performed constraining the 
$\psi$ mass to its nominal value as given by the PDG~\cite{PDG2024} and requiring the $ B^+ $ momentum vector to point back to its primary vertex. 
Candidates satisfying
these criteria
are retained for further consideration.

Background is suppressed using two BDT classifiers~\cite{Breiman,AdaBoost} implemented in the TMVA 
toolkit~\cite{Hocker:2007ht,*TMVA4}.
The first classifier 
is trained using fully simulated $B^+ \to \psi (2S) \phi K^+$ samples as a proxy for the signal and events from the far-upper sideband 
($5.8 < M(\psi(2S)\phi K^+) < 6.0\GeVcc$) as a proxy for the background. 
The variables with highest discrimination power are used as input to the classifier.
There are 21 variables that include: 
the \pt and \chisqip of the kaons, muons, $\psi$, $\phi$, 
and the  $ B^+ $ candidates, the cosine of the angle between the  $ B^+ $ momentum and the vector connecting the 
PV and the $ B^+ $ decay vertex, 
the sum of the \chisqip of all final-state 
particles with respect to the PV, the vertex $\chi^2$ of the $B^+$ candidate, and
the $ B^+ $ decay time and its significance. 
The chosen BDT threshold  corresponds to a signal efficiency close to $ 90\%$ 
and a background rejection of about $99\%$.

After all previous steps,  more than one 
\BtopsiphiKDecay candidate 
in a single event can satisfy
the selection criteria.
In almost all of these cases (around 30\% and 3\% of events in the signal and normalization channels, respectively), this occurs when the masses of the two
$ K^+ K^- $ combinations in a single $ B^+ \to \psi K^+ K^- K^+ $
candidate fall within the
nominal $ \phi $ window. 
Combinatorial background may also produce additional candidates in a single event, though on a smaller scale.
To retain one $B^+$ candidate per event,
a second BDT is trained with five variables as input.
The input variables include properties of the $B^+ $ meson,  such as the \chisqip, the decay-time significance, the decay-vertex quality, the cosine of the angle between its momentum vector and the trajectory. Additionally, the input variables include the significance of the maximum distance between the trajectories of final-state particles.
The candidate with the highest BDT
output in each event is selected and its
mass is used in the $B^+$ mass fit
described below.
The correlation between the mass of the $B^+$ candidate 
and the BDT output is found to be negligible. 
When the masses of both $ K^+ K^- $ pairs 
in a $ B^+ \to \psi K^+ K^- K^+ $ candidate
fall within the 
nominal $ \phi $ window,
both $K^+ K^-$ combinations
are used in the fit to determine $ F_{\rm Signal} \ (F_{\rm Norm})$.

\subsection{Mass fits}
\label{sec:MassFits}

\begin{figure}
\begin{center}
    \includegraphics*[width=.49\textwidth]{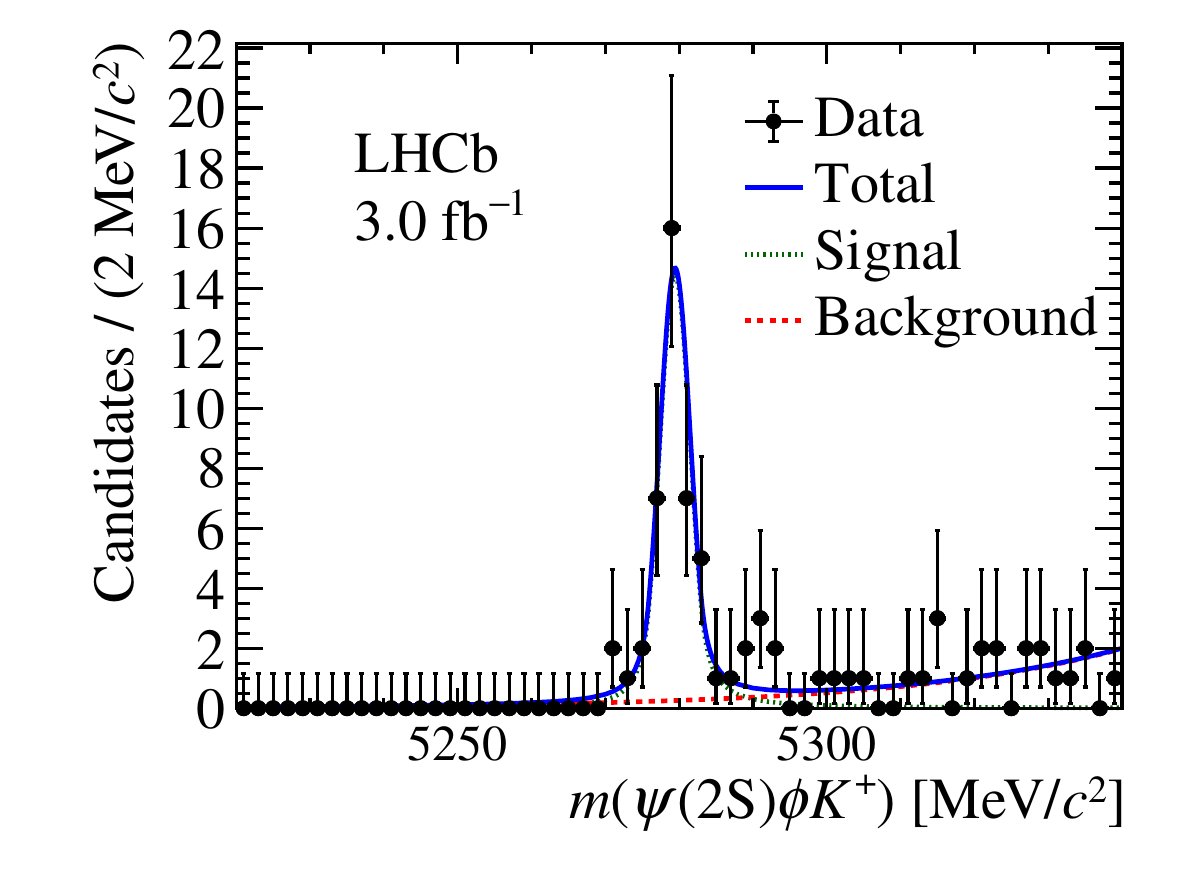}
    \includegraphics*[width=.49\textwidth]{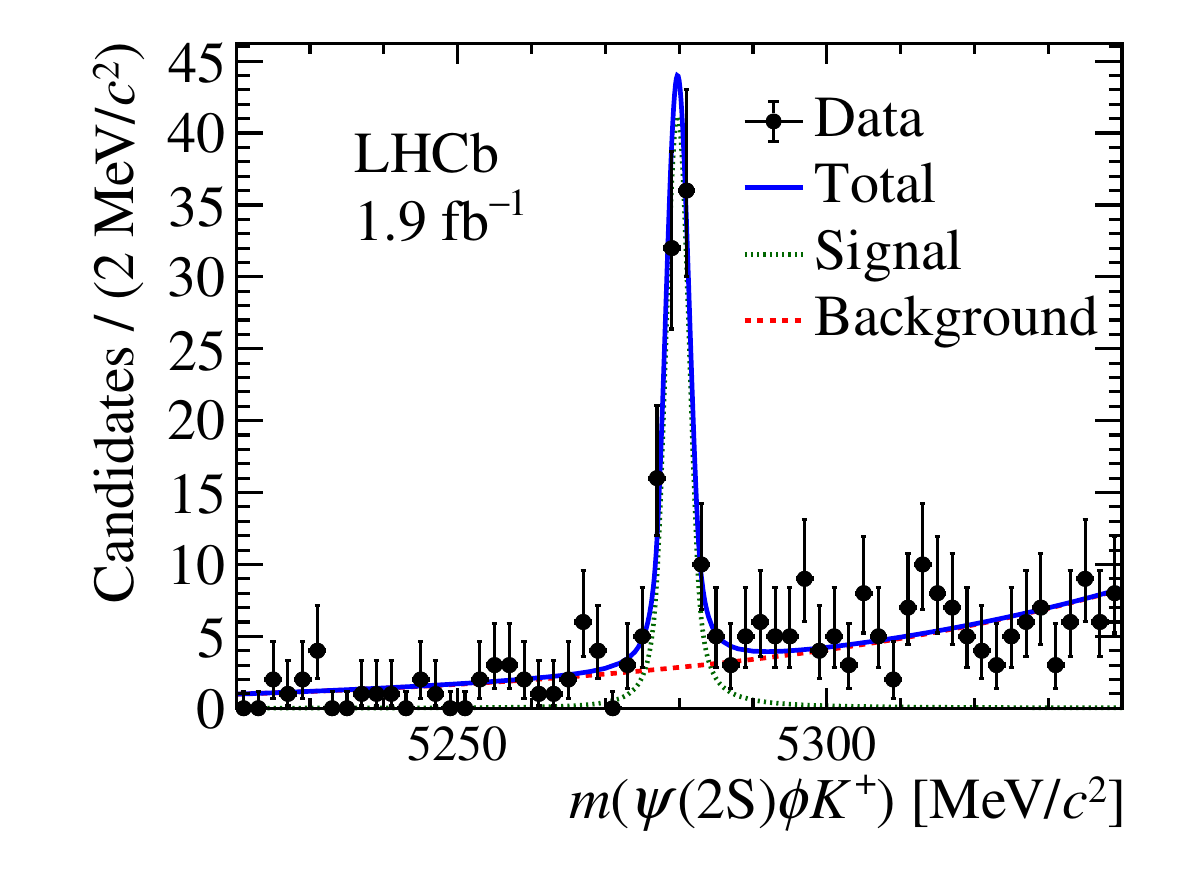}
    \includegraphics*[width=.49\textwidth]{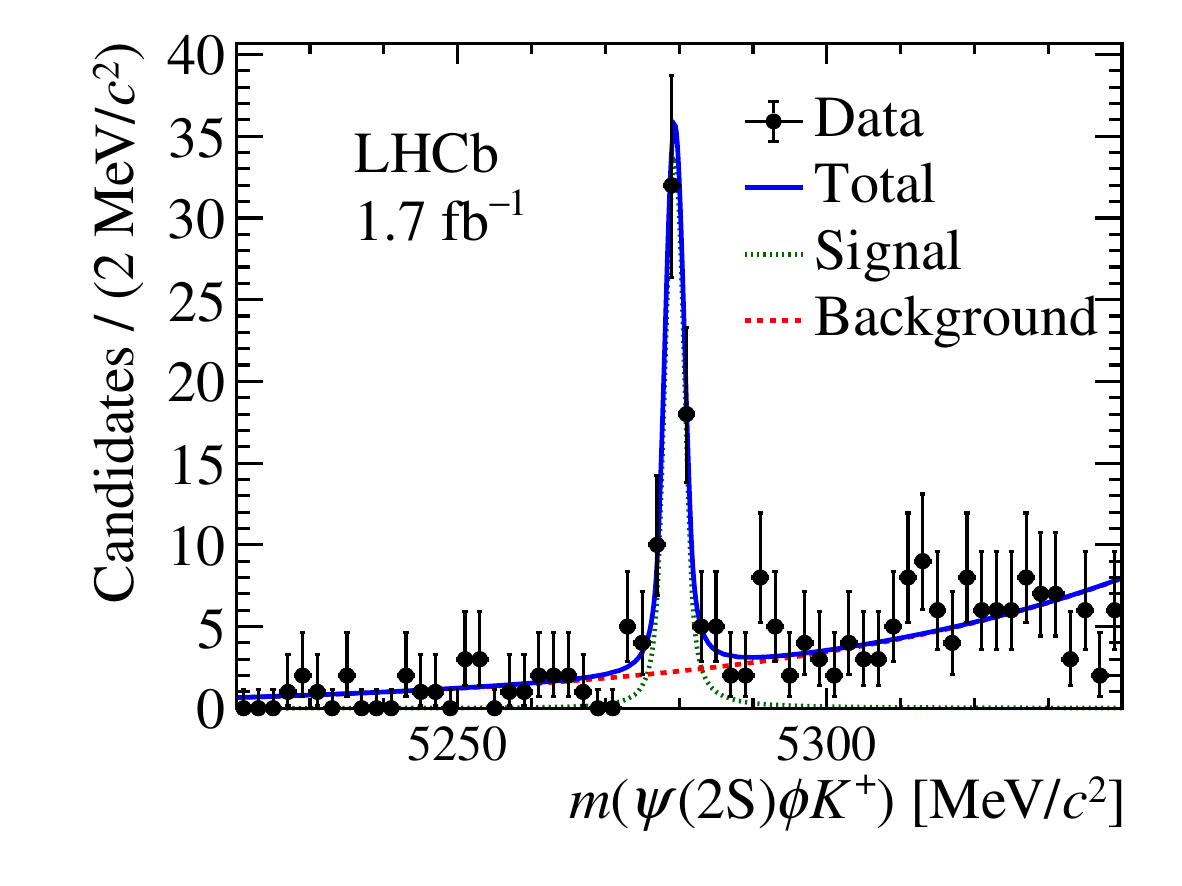}
    \includegraphics*[width=.49\textwidth]{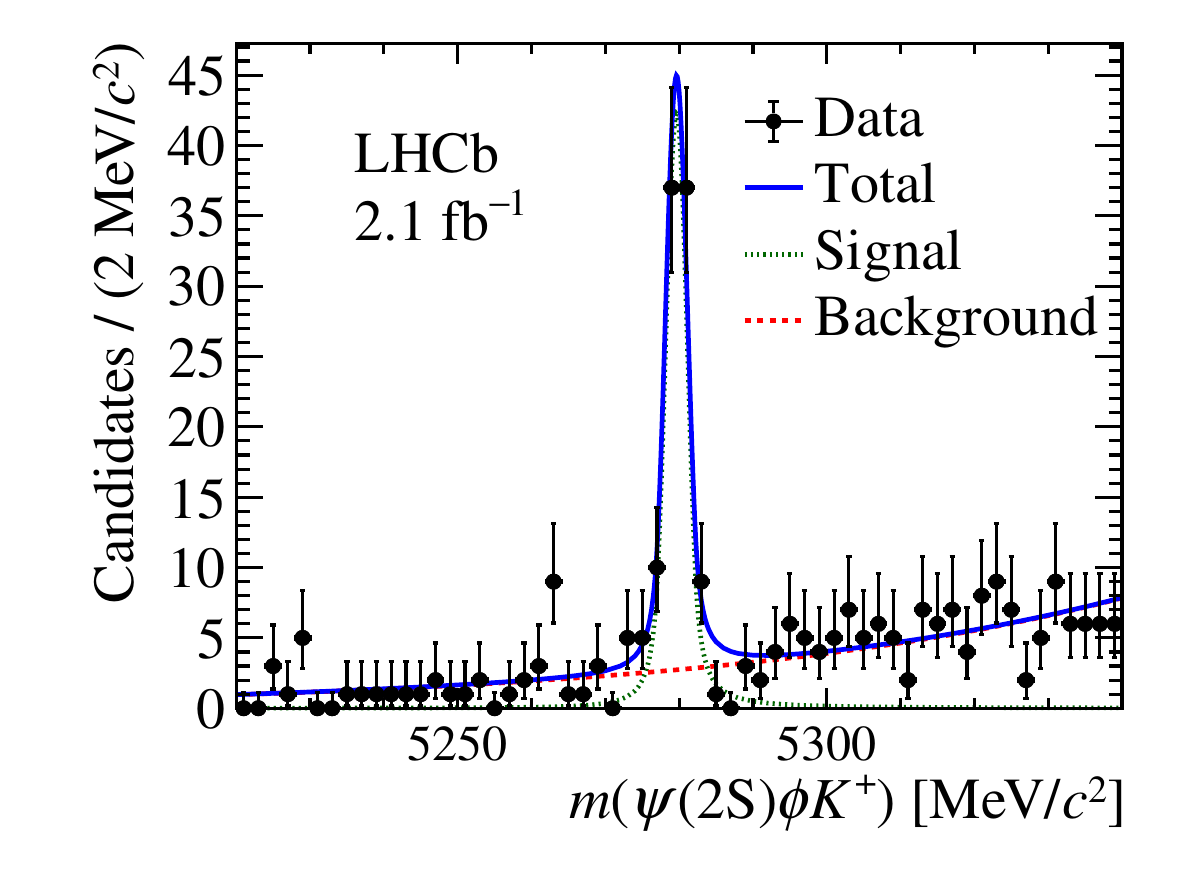}
\end{center}
\vspace*{-.7cm}
\caption{\small
Invariant-mass distributions of $B^+\rightarrow \psi(2S) \phi K^+$ 
candidates with fit projections for 
(top left) Run 1, (top right) 2015--2016, (bottom left) 2017, and (bottom right) 2018  samples. }
\label{fig:SignalMassFit}
\end{figure}

\begin{figure}
\begin{center}
    \includegraphics*[width=.49\textwidth]{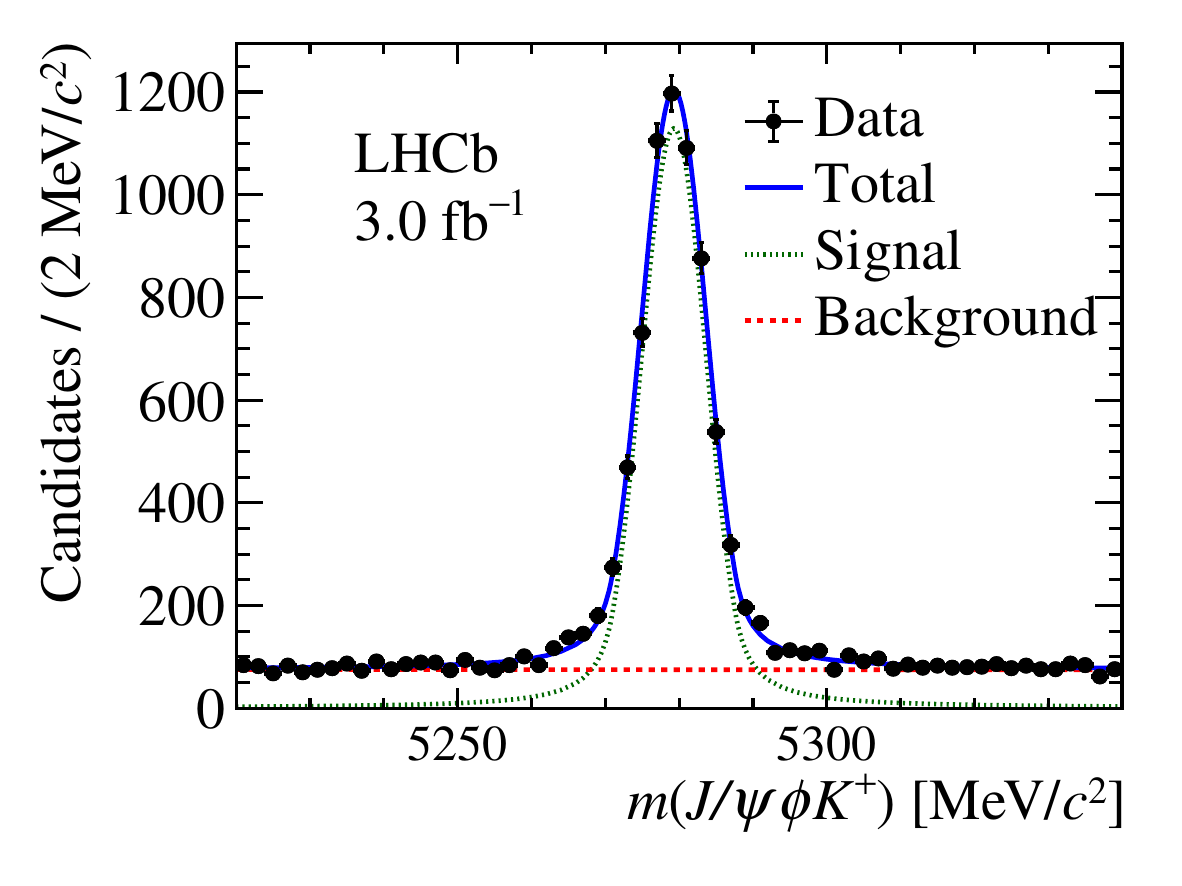}
    \includegraphics*[width=.49\textwidth]{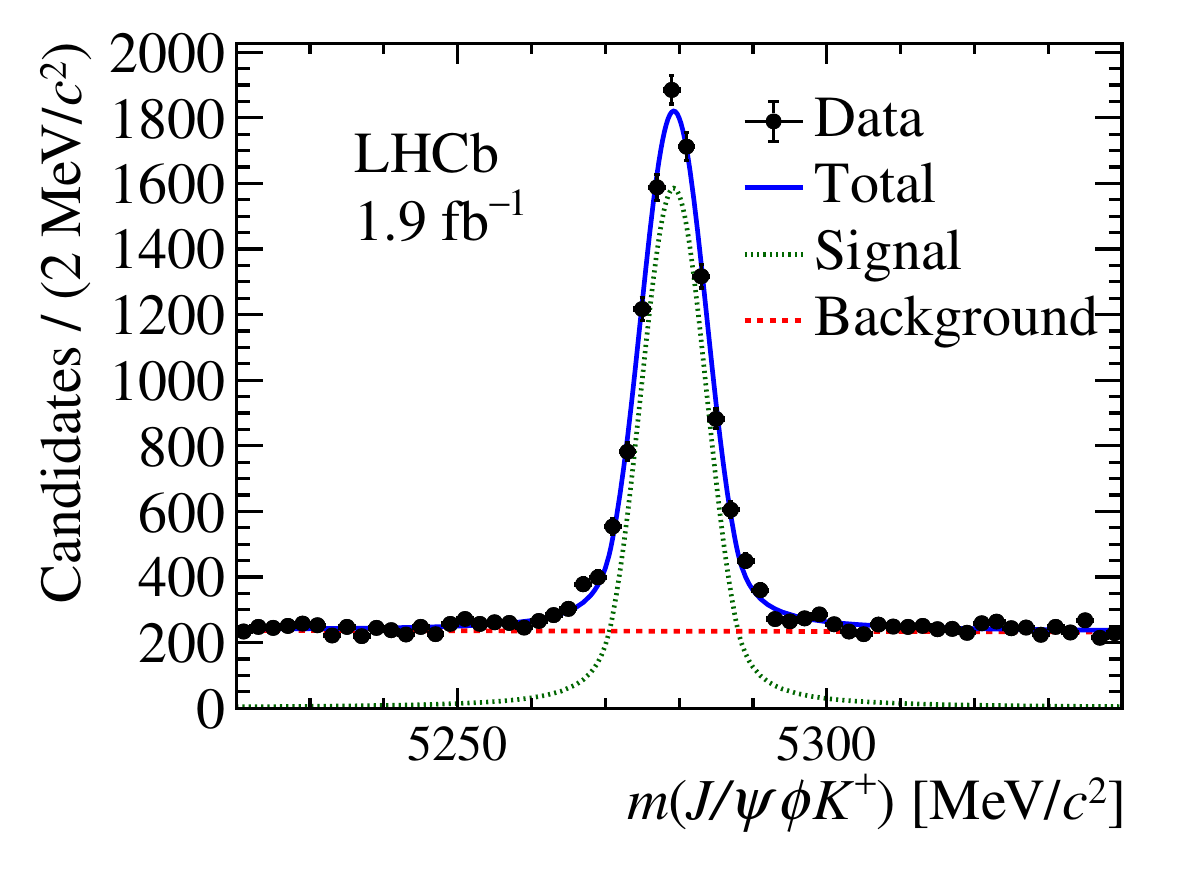}
    \includegraphics*[width=.49\textwidth]{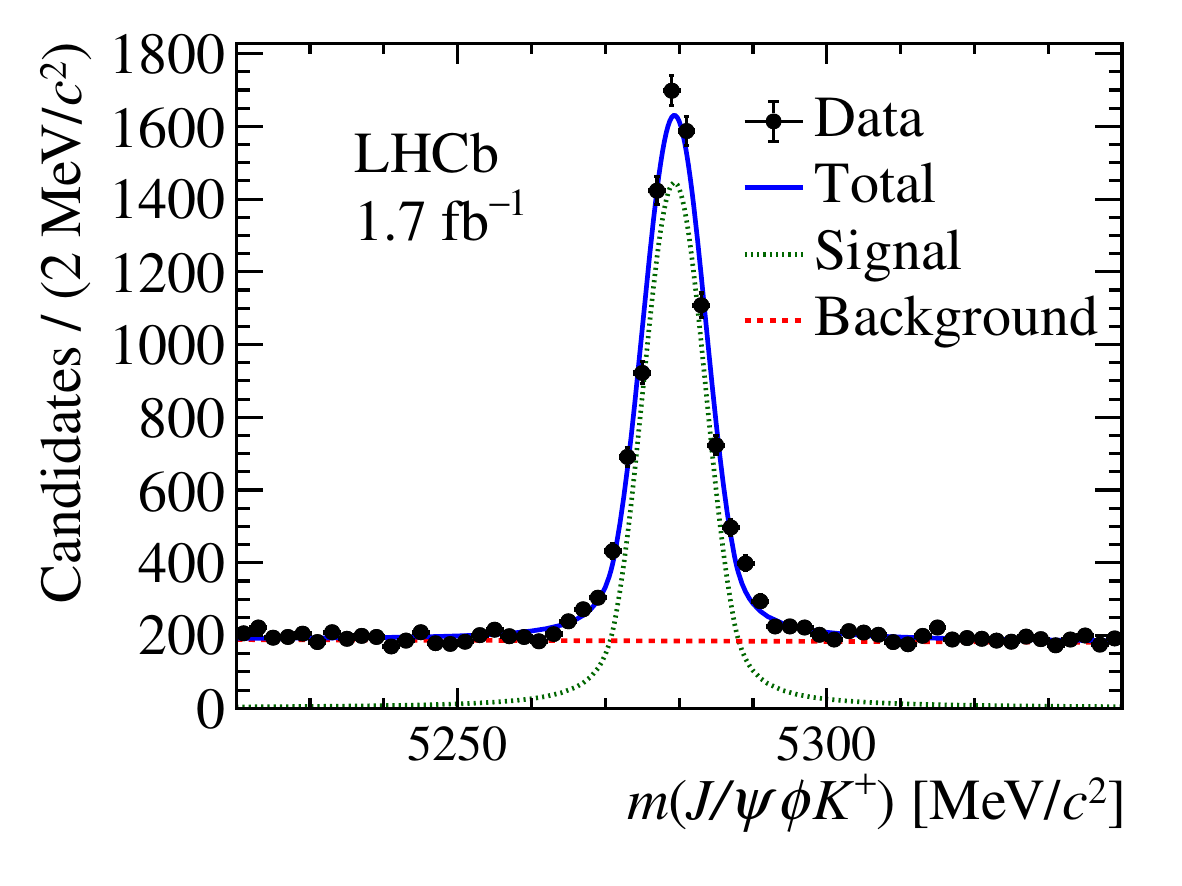}
    \includegraphics*[width=.49\textwidth]{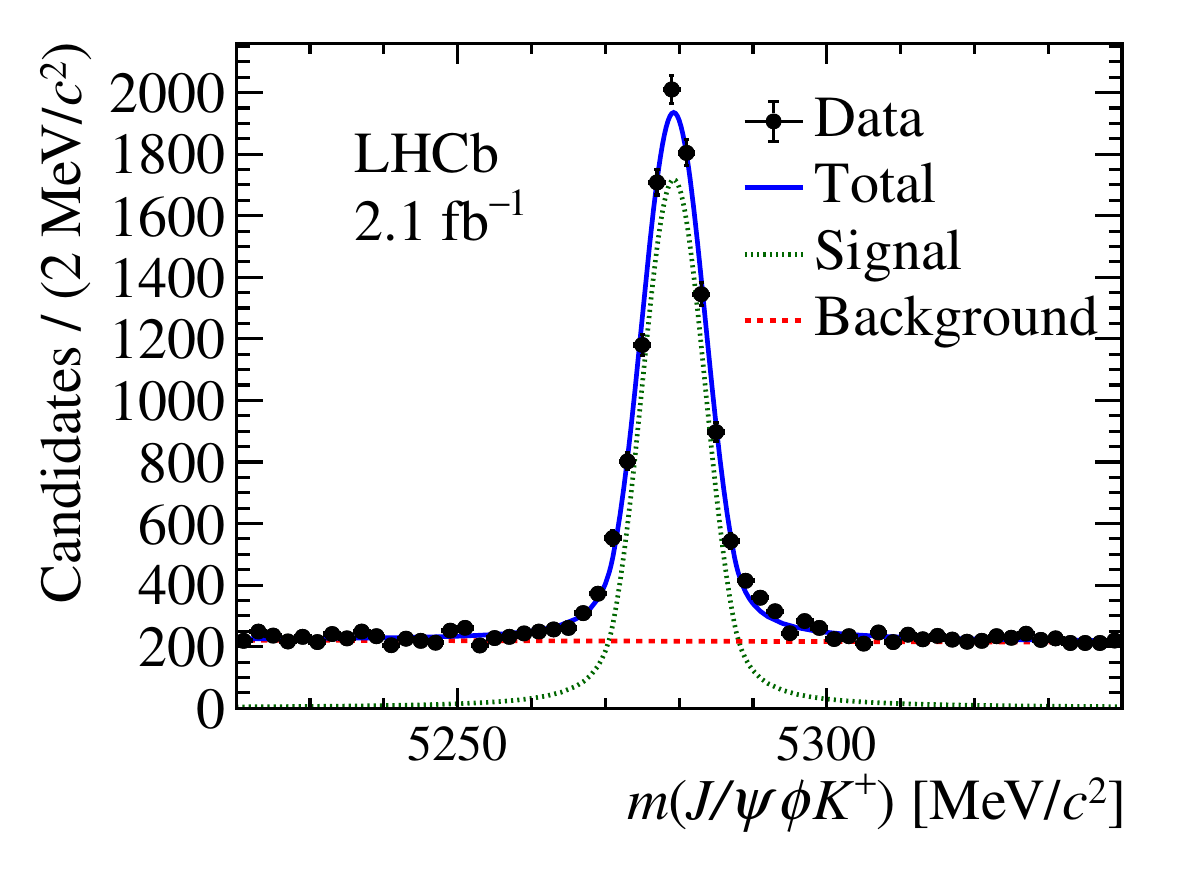}
\end{center}
\vspace*{-.7cm}
\caption{\small
Invariant-mass distribution of $B^+\rightarrow J/\psi \phi K^+$ candidates with fit projections for 
(top left) Run~1, 
(top right) 2015--2016, 
(bottom left) 2017, and 
(bottom right) 2018  samples. }
\label{fig:ControlMassFit}
\end{figure}

Unbinned maximum-likelihood mass fits are performed separately for the \BtopsiphiKDecay and \BtoJpsiphiKDecay samples, and for each data-taking period as the detector operating-conditions
varied for each interval. 
For both the signal and normalization 
channels, the $B^+$ peak 
is modeled by a double-sided Crystal Ball  
function (DSCB)~\cite{Skwarnicki:1986xj}, while the background is described by an exponential function. 
The DSCB tail parameters are obtained from simulation
and are fixed in the fits to the data. 
The yields of signal and background, 
the signal peak position $M_{B^+}$ and width $\sigma_{B^+}$, and the parameters describing the exponential backgrounds,
are floated in the fits.
Mass distributions of \BtopsiphiKDecay
and \BtoJpsiphiKDecay candidates 
are shown in Figs. \ref{fig:SignalMassFit} and \ref{fig:ControlMassFit}, respectively, along
with the results of the fits. 
The fit parameters of each data taking periods are reported in Table~\ref{tab:MassFit}. The yields of the signal and normalization models determined from the fits over the combined samples
are $289\pm 19$ and $35\,315\pm 237$ events, respectively, which are minimally
compatible with the sum of the yields obtained independently, providing additional confidence in the consistency of these values.

\begin{table}
	\caption{\small 
Results of the fit to the invariant mass of signal and normalization decays.}
	\begin{center}
    \begin{tabular}{|c|c| r@{\:$\pm$\:}l | r@{\:$\pm$\:}l| }\hline
		 Period & Parameter & \multicolumn{2}{c|}{$B^+ \rightarrow \psi(2S) \phi K^+$} & \multicolumn{2}{c|}{$B^+ \rightarrow \jpsi \phi K^+$} \\ \hline
		{} & $M_{B^+}$ $[\mevcc]$ & $ 5279.45  $ & $ 0.40 $ & $ 5279.30  $ & $ 0.07 $\\
		  Run 1 & $\sigma_{B^+}$ $[\mevcc]$ & $ 2.04  $ & $ 0.41 $ & $ 4.38  $ & $ 0.07 $\\
		 {} & $N_{\rm{Signal(Norm)}}$ & $ 43  $ & $ 7 $ & $ 6835  $ & $ 97 $\\
		\hline
		{} & $M_{B^+}$ $[\mevcc]$ & $ 5279.79  $ & $ 0.22 $ & $ 5279.28  $ & $ 0.06 $\\
		{2015--2016}  & $\sigma_{B^+}$ $[\mevcc]$ & $ 1.60  $ & $ 0.23 $ & $ 4.46  $ & $ 0.07 $\\
		 & $N_{\rm{Signal(Norm)}}$ & $ 95  $ & $ 11 $ & $ 9772  $ & $ 129 $\\
		\hline
		{}& $M_{B^+}$ $[\mevcc]$ & $ 5279.40  $ & $ 0.22 $ & $ 5279.36  $ & $ 0.06 $\\
		  {2017}  & $\sigma_{B^+}$ $[\mevcc]$ & $ 1.33  $ & $ 0.23 $ & $ 4.26  $ & $ 0.07 $\\
		 {} & $N_{\rm{Signal(Norm)}}$ & $ 64  $ & $ 9 $ & $ 8515  $ & $ 118 $\\
		\hline
		{}  & $M_{B^+}$ $[\mevcc]$ & $ 5279.65  $ & $ 0.19 $ & $ 5279.27  $ & $ 0.06 $\\
		  {2018} & $\sigma_{B^+}$ $[\mevcc]$ & $ 1.48  $ & $ 0.16 $ & $ 4.29  $ & $ 0.06 $\\
		  {} & $N_{\rm{Signal(Norm)}}$ & $ 90  $ & $ 11 $ & $ 10196  $ & $ 128 $\\
		\hline
		  \end{tabular}\end{center}
	\label{tab:MassFit}
\end{table}

\begin{figure}
\begin{center}
    \includegraphics*[width=.49\textwidth]{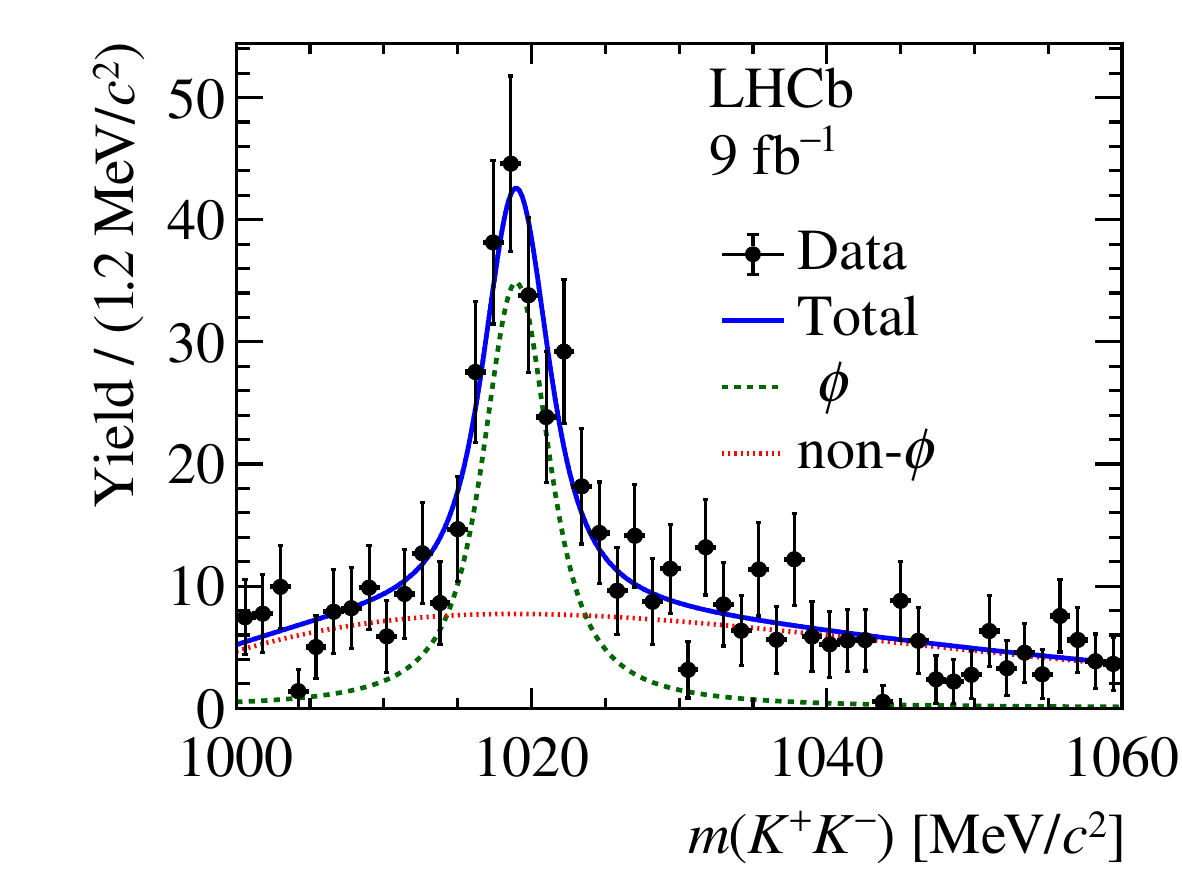}
    \includegraphics*[width=.49\textwidth]{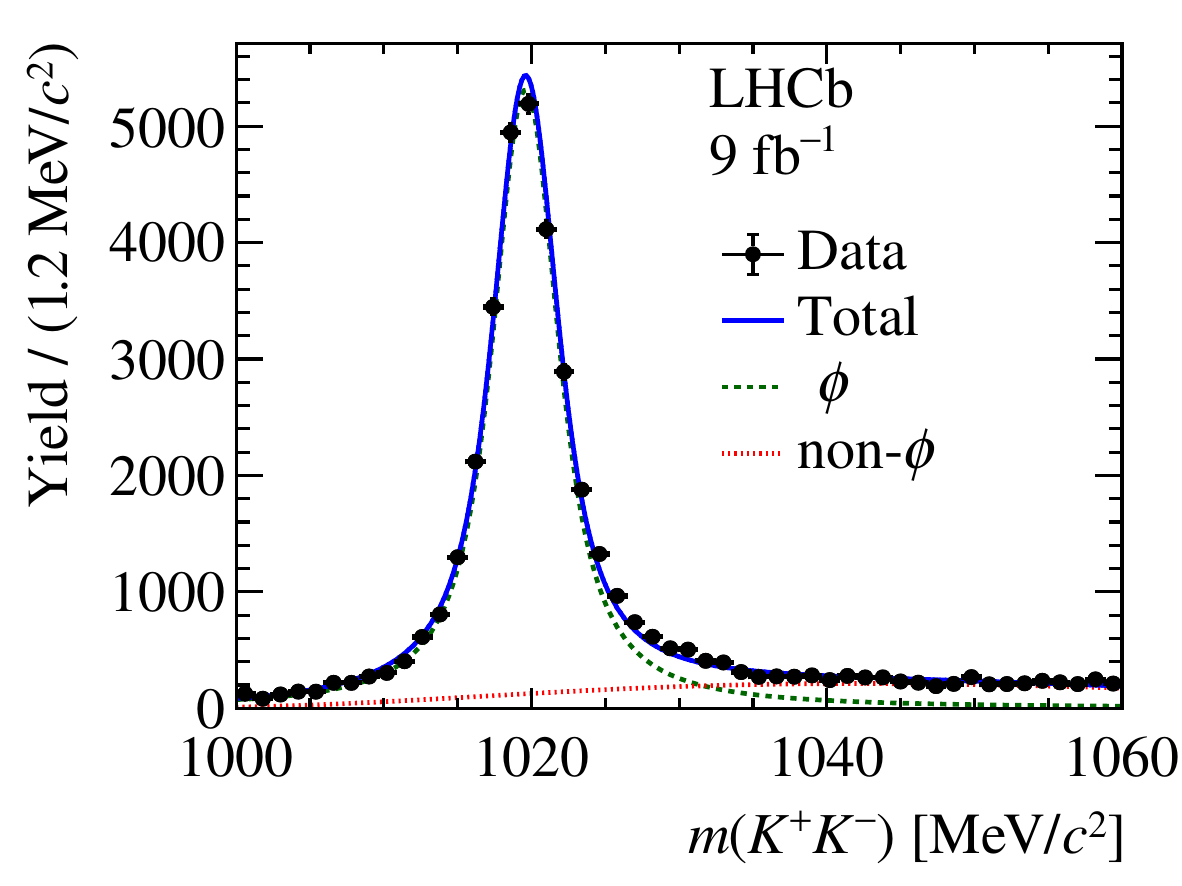}
\end{center}
\vspace*{-.7cm}
\caption{\small
Distributions of the  background-subtracted $K^+K^-$  mass spectra from the full (Run 1 + Run 2) data sample for (left) $B^+ \to \psi(2S) \phi K^+$ and (right) $B^+ \to J/\psi \phi K^+$ decays.}
\label{fig:Phi}
\end{figure}

Although the samples are selected with $ K^+ K^-$ invariant masses in 
the $\mbox{1000--1040\mevcc}$ mass window, 
significant non-$ \phi $ contributions are present. 
The fraction of $\phi$ candidates is obtained from a binned $\chi^2$ fit to the extended $K^+ K^-$ mass distribution of 1000--1060$\mevcc$.
The selection criteria are the same as for the \Bp samples
except for this relaxed  $K^+K^-$ mass range.
In these fits, background is statistically subtracted 
by means of 
the \sPlot technique~\cite{sWeights}, 
using the results of the fit to the $B^+$ mass distributions. 
The mass distribution of the $\phi$ component 
is described by a P-wave Breit--Wigner function, convolved with a Gaussian function with resolution fixed to 1\mevcc (taken from simulation), while that of the non-$\phi$ component is described by an Argus function~\cite{ARGUS:1990hfq}
whose parameters are floated. 
As the yields of signal in the disjoint data samples are small while the event-selection efficiency does not vary much in the $K^+K^-$ mass window, 
both $B^+$ and $K^+K^-$ mass fits are performed over the integrated Run~1 + Run~2  data samples. 
Therefore, only one $F_{{\mathrm{Signal(Norm)}}}$ is calculated for each 
of the signal and normalization channels and used for all
disjoint data samples.
The $K^+K^-$ mass fit results are shown in Fig. \ref{fig:Phi}, 
in which the different line shapes of the non-$\phi$ component are mainly due to the different phase space of the signal and normalization modes. 
The fractions of $\phi$ contribution
within the nominal $K^+K^-$ mass window are 
determined to be 
$F_{{\mathrm{Signal}}} = 0.73\pm 0.10$
and $F_{{\mathrm{Norm}}} = 0.91\pm 0.01$, 
where the uncertainties are statistical only. 
The values of $ \chi^2 $ per degree  of freedom for the
signal and normalization fits are $ \chisqndf = 1.2 $ and
$ \chisqndf = 2.3 $, respectively, both with the
number of degrees of freedom $\mathrm{ndf}=95$.

\subsection{Efficiencies}
\label{sec:Efficiency}
Particle-identification and tracking efficiencies 
are obtained from calibration data samples~\cite{Aaij:2018vrk}.
The remaining efficiencies, for both the signal and normalization decay modes,
are obtained from simulated samples and 
calculated separately for each data-taking period. 
The normalization simulation sample, generated as pure phase space,
is weighted to account for the structures in the $\psi \phi$ and $\psi K^+$ masses observed in data.
No similar correction has been applied to the signal
simulation sample as no structure is visible in the Dalitz plot of the signal data sample.
A systematic uncertainty is estimated to address possible residual effects.
In the normalization sample, per-candidate weights are calculated by training a gradient-boosted decision-tree reweighter, GBReweighter, implemented in \texttt{hep\_ml}~\cite{hep_ml}. 
The data distributions used here have the background subtracted statistically 
by means of the \sPlot technique~\cite{sWeights},
using the results of the fit to the  \BtoJpsiphiKDecay
invariant-mass distribution.

Two factors distort the distribution of events in phase space (the $\psi,\phi,K^+ $ Dalitz plot), 
especially at the edges. 
First, the resolution of the reconstructed $B^+$ invariant mass cannot be neglected.
To mitigate this problem, the masses of the $B^+$  and $\psi$ candidates are constrained to their nominal values. 
Second, the natural line width of the $\phi$ meson cannot be neglected. 
Therefore,  the simulated samples are reweighted in the
$(s_{\psi K}, \cos(\theta))$ space, where
$s_{XY} $ denotes the invariant-mass squared of the $ XY $ system and
$\theta $ is the angle between the $\psi$ and $\phi$, in the $\phi K^+$ rest frame, calculated as
\begin{equation}
    \cos(\theta) = \frac{(m_{B^+}^2 - s_{\phi K} - m_{\psi}^2)( s_{\phi K} + m_{\phi}^2 - m_{K}^2) + 2 s_{\phi K}(m_{\psi}^2 + m_{\phi}^2 - s_{\psi\phi})}
                    {\lambda^{\frac{1}{2}}(m_{B^+}^2,s_{\phi K}, m_{\psi}) \,\lambda^{\frac{1}{2}}(s_{\phi K},  m_{\phi}^2, m_{K}^2 ) } \, ,
\end{equation}
\noindent
where $m_{B^+}, m_\psi$ and $m_\phi$ are the reconstructed 
mass of the $B^+$, $\psi$ and $\phi$ mesons, respectively, $m_K$ is the known $K^+$ mass~\cite{PDG2024}, and 
$\lambda^{\frac{1}{2}}(x,y,z)$ is defined as

\begin{equation}
\lambda^{\frac{1}{2}}(x,y,z) \equiv \sqrt{(x - y - z)(x - y - z) - 4yz} \, .   
\end{equation}
As this parametrization is not directly correlated with the $\phi$ mass, 
and $ -1 < \cos(\theta)<1$ by definition, the weighting at the 
limits of phase space is better behaved than if the usual Dalitz plot variables are used.

Ratios of efficiencies for the signal and normalization channels are
derived from weighted, simulated data for each running period and
are reported in Table~\ref{tab:Efficiencies}.
Although the hardware and first-stage software triggers are more efficient for signal channel  decays,
the second-stage software trigger and offline selection criteria are more efficient for the
normalization channel decays.
This results from the much smaller phase space for the kaons in the signal channel
and the use of the track $\chi^2_{\rm IP}$
to exclude candidates with kaon momentum and \Bp flight directions consistent with each other.

\begin{table}
  \caption{\small 
  Ratios of efficiencies 
  for the signal and normalization channels 
  for each data-taking period,
  as described in the text, where the uncertainties are statistical only. 
  }
    \begin{center}\begin{tabular}{|c|c|}\hline
      Period    & $\epsilon_{\rm Signal}/\epsilon_{\rm Norm}$     \\ \hline
      Run 1     & $0.823 \pm 0.004$ \\ 
      2015--2016 & $0.801 \pm 0.007$ \\ 
      2017      & $0.790 \pm 0.004$ \\ 
      2018      & $0.834 \pm 0.003$ \\
      \hline 
    \end{tabular}
  \end{center}
  \label{tab:Efficiencies}
\end{table}

\subsection{Systematic uncertainties}
\label{sec:Systematics}

Six sources of systematic uncertainty are discussed.
Others were considered, but judged to be negligible
compared to the statistical and overall systematic
uncertainties reported.

The model used to describe the signal and background shapes may affect the results. 
Where the baseline fit uses a DSCB function to model the signal, a Johnson $S_{U}$ function~\cite{10.1093/biomet/36.1-2.149} is used as an alternative model to fit the data. 
Similarly, the baseline exponential background shape is replaced with a second order Chebychev polynomial.
Using the full dataset, results obtained with Johnson $S_{U}$ show a difference of $\sim 2\%$ in both signal and normalization samples. 
Added in quadrature, the total systematic uncertainty associated with the signal and background shapes is $\sim 3\%$.

The fractions of $\phi$ contributions
in the nominal $ m(K^+K^-) $ signal windows, $F_{\rm Signal} $ and $F_{\rm Norm} $,  are also  subject to fluctuation. 
Their
statistical uncertainties
are propagated to the branching fraction ratio measurements.
These lead to
a systematic effect of around 14\%, almost entirely associated with
the signal channel. 
This is the dominant systematic uncertainty.
Increasing the uncertainty associated with $F_{\rm Norm} $ to account for the
large $ \chisqndf$ value associated with its fit projection
does not change the uncertainty in the $ \phi $ fraction ratio
at the level of precision reported.

The Breit--Wigner function used to model the $ \phi$ mass distribution is not corrected for the $B^+ \to \psi K^- K^+ K^+$ four-body  phase-space shape. 
Fitting the reweighted simulation samples with the same procedure yields a 3\% deviation in the ratio of the fraction
of  $\phi$ events between signal and normalization samples. 

The limited  sizes of simulation samples lead to statistical uncertainties in the efficiencies. 
These are propagated to the final results as systematic uncertainties,
explicitly accounting for the weighting procedure discussed below.
The relative uncertainty from this source varies
between 3\% and 4\%. 

The weighting procedure accounting for resonant structure in phase space requires
a set of four parameters as input to 
train the gradient-boosted decision-tree reweighter~\cite{hep_ml}.
On trialing a large number of alternative parameter sets, the distribution of results shows a negligible spread.
Therefore, no systematic uncertainty is associated with weighting the normalization channel. 
As the weighting procedure is not applied to the signal model in the baseline fit,
the associated 
systematic uncertainty
is estimated by taking the difference between the efficiency with and without this weighting. 
Uncertainties of 2\%, 1\%, 3\% and 4\% are found for the Run~1, 2015--2016, 2017 and 2018 samples, respectively.

The  branching fractions
$\BR(\Jpsitomumu)$ and $\BR(\psitomumu)$ are required 
to calculate the relative branching fraction $\BR(\BtopsiphiKDecay)/\BR(\BtoJpsiphiKDecay)$, 
and  the branching fraction of the $\BtoJpsiphiKDecay$ is additionally required for the  $\BR(\BtopsiphiKDecay)$ calculation. 
These values are taken from the PDG~\cite{PDG2024} and 
their uncertainties are propagated to the final results. 
The contribution to the systematic uncertainty of the branching fraction ratio is $ \sim 2\%$  and the corresponding contribution to the \BtopsiphiKDecay branching fraction is $ \sim 8\%$.

Table \ref{tab:Syst} summarizes the systematic uncertainties. With the exception of the uncertainties due to simulation sample sizes and signal weighting, systematic uncertainties are assumed to be fully correlated throughout all data-taking periods. 

\begin{table}
  \caption{\small Systematic uncertainties on the ratio of branching fractions of the decays 
  $\BtopsiphiKDecay$ and $\BtoJpsiphiKDecay$, by origin and data-taking period.}
  \begin{center}
\resizebox{1\columnwidth}{!}{
\begin{tabular}{|c|c|c|c|c|c|c|c|}\hline
      Period    & \begin{tabular}[c]{@{}c@{}} Mass \\ model \end{tabular}  & $\phi$ fraction  & \begin{tabular}[c]{@{}c@{}} Phase-space \\ correction \end{tabular}  & \begin{tabular}[c]{@{}c@{}} Simulation \\ statistics \end{tabular} 
      & \begin{tabular}[c]{@{}c@{}} Signal\\ efficiency \\ correction \end{tabular}  & \begin{tabular}[c]{@{}c@{}}  External \\
      inputs~\cite{PDG2024}  \end{tabular} & \begin{tabular}[c]{@{}c@{}}  Total \\
        \end{tabular} \\ \hline

      Run 1     & 3\% & 14\% & 3\% & 4\% & 2\% & 2\%  & 15\%\\                                
      2015--2016 & 3\% & 14\% & 3\% & 3\% & 1\% & 2\%  & 15\%\\                                
      2017      & 3\% & 14\% & 3\% & 3\% & 3\% & 2\%  & 15\%\\                                
      2018      & 3\% & 14\% & 3\% & 3\% & 4\% & 2\%  & 15\%\\                                
      Combined  & 3\% & 14\% & 3\% & 1\% & 1\% & 2\%  & 15\% \\                                
      \hline 
    \end{tabular}}
    \end{center}
\label{tab:Syst}
\end{table}

\section{Results and summary}
\begin{figure}
\begin{center}
    \includegraphics*[width=.8\textwidth]{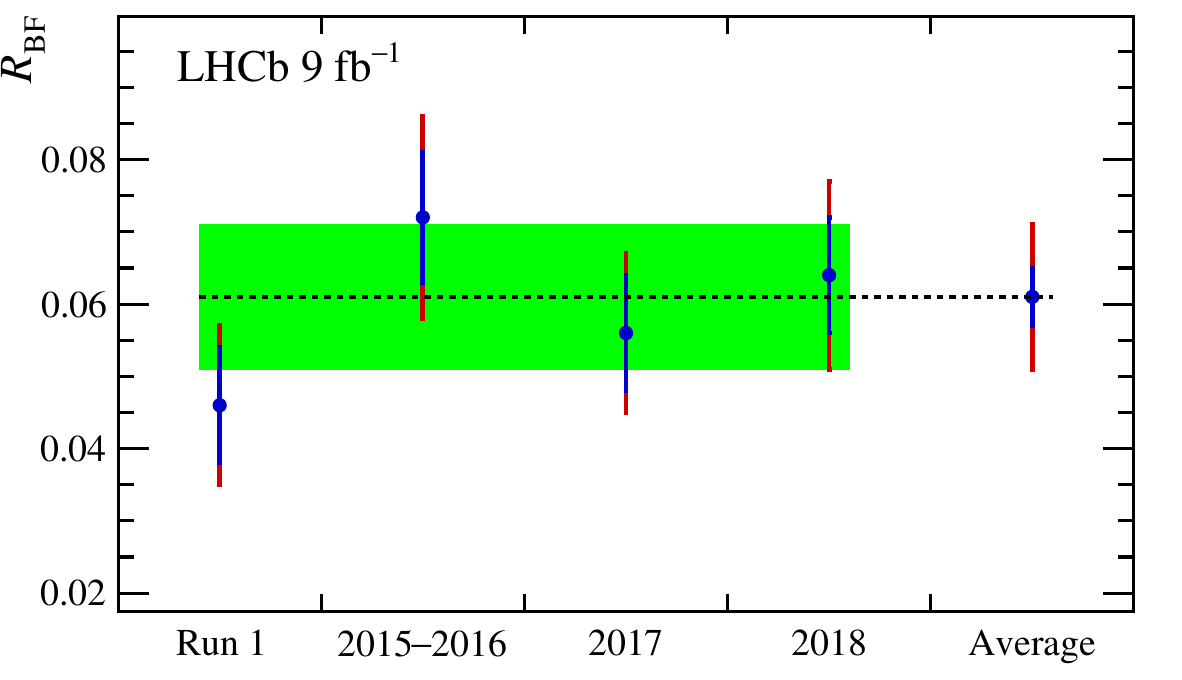}
\end{center}
\caption{\small
Ratio $\mathcal{R}_{\rm BF}$ of branching fractions for the decays $B^+ \to \psi(2S) \phi K^+$  and $ B^+ \to J/\psi \phi K^+$ for each data-taking period and 
the confidence level region at one standard deviation for the combined result in green. 
The inner error bars (in blue) and the outer error bars (in red) delimit the statistical and total (statistical plus systematic, added  in quadrature) uncertainties, respectively. }
\label{fig:RelBF}
\end{figure}

\begin{figure}
\begin{center}
    \includegraphics*[width=.9\textwidth]{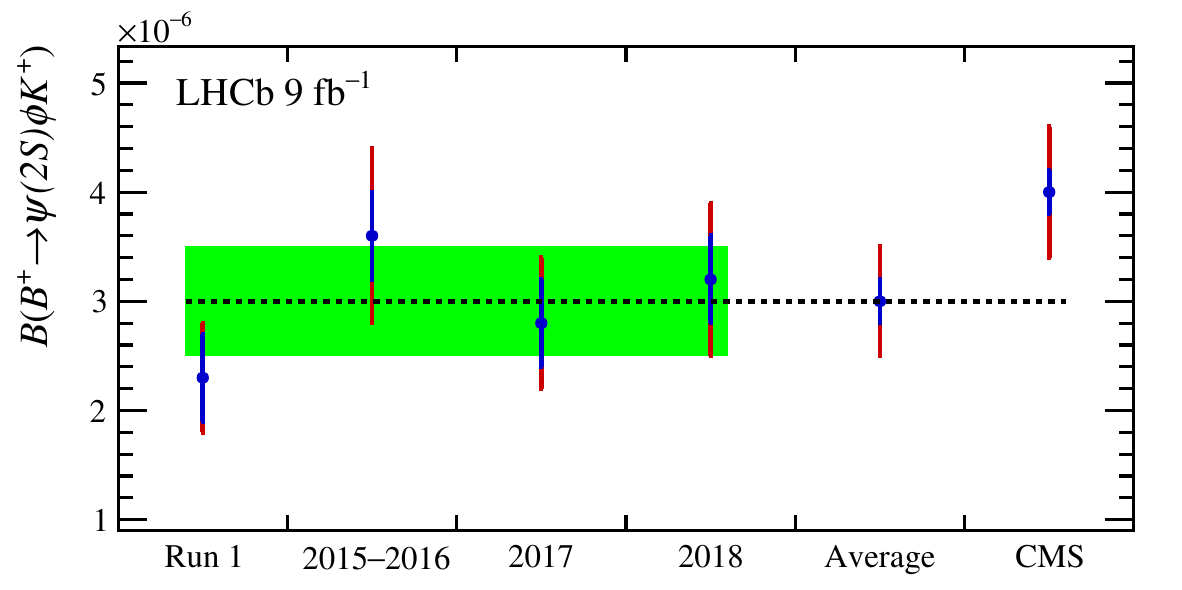}
\end{center}
\caption{\small
Branching fraction of the decay $B^+ \to \psi(2S) \phi K^+$ for each
data-taking period and the confidence level region at one standard deviation for the combined result in green. 
The inner error bars (in blue) and the outer error bars (in red) delimit the statistical and total (statistical plus systematic, added in quadrature) uncertainties, respectively. 
The CMS result~\cite{Khachatryan:2016npv} is included for comparison.}
\label{fig:AbsBR}
\end{figure}

Using the yields, the $\phi $ fractions and efficiency ratios reported in Tables~\ref{tab:MassFit} and \ref{tab:Efficiencies}, respectively, 
the ratio of branching fractions between the \BtopsiphiKDecay and \BtoJpsiphiKDecay 
decays is calculated. 
The results are reported in Table \ref{tab:RelBF}
and illustrated in Fig. \ref{fig:RelBF}.
The derived \BtopsiphiKDecay  branching fraction results are illustrated in Fig.~\ref{fig:AbsBR}.
Results for each data-taking period are combined using the Best Linear Unbiased Estimator method~\textsc{Blue}~\cite{Nisius:2020jmf}, which accounts for correlations among systematic effects.
The reported results are consistent throughout the data-taking periods.

The combined value obtained for the ratio of branching fractions is
\begin{displaymath}
\mathcal{R}_{\rm BF}\equiv
  \frac{\BR(\BtopsiphiKDecay)}{\BR(\BtoJpsiphiKDecay)} = 0.061 \pm 0.004 \pm 0.009 \, ,
\end{displaymath}
where the first uncertainty is statistical and the second is systematic.
The corresponding value for 
the \BtopsiphiKDecay branching fraction is 
\begin{equation}
\BR(\BtopsiphiKDecay) = (3.0 \pm 0.2 \pm 0.5 \pm 0.2) \times 10^{-6} \nonumber \, ,
\end{equation}
where the third uncertainty is associated with the uncertainty of the branching fraction of the normalization channel. 
The result is compatible with the CMS measurement 
$(4.0 \pm 0.4 \pm 0.6 \pm 0.2) \times 10^{-6}$~\cite{Khachatryan:2016npv}.

The phase space in \BtopsiphiKDecay decays is about 80 times 
smaller than that in \BtoJpsiphiKDecay decays.
Considering 
Fermi's golden rule that the transition rate is 
proportional to the phase space times the magnitude of the amplitude squared, 
one may conclude that the average amplitude squared for the \BtopsiphiKDecay decay is approximately 5 times larger than 
that for the \BtoJpsiphiKDecay decay.
This suggests the presence of resonant amplitudes in the 
\BtopsiphiKDecay decay that enhance its average decay rate.
States such as the $\chi_c(4700)$~\cite{Dias:2013qga,Voloshin:2019ilw}
have the right masses and quark content to play this role,
so an amplitude analysis made possible with an increased data sample size
represents a promising avenue for further investigation.

\begin{table}
  \caption{\small 
  Ratio of branching fractions ${\mathcal{R}}_{\rm BF}$ 
  for each data-taking period and for the combined sample, where the first uncertainty is statistical and the second is systematic.}
    \begin{center}\begin{tabular}{|c|c|}\hline
      Period    & ${\mathcal{R}}_{\rm BF}$   \\ \hline
      Run 1     & $0.046 \pm 0.008 \pm 0.007$    \\ 
      2015--2016 & $0.072 \pm 0.009 \pm 0.011$    \\ 
      2017      & $0.056 \pm 0.008 \pm 0.008$    \\ 
      2018      & $0.064 \pm 0.008 \pm 0.010$    \\
      Combined  & $0.061 \pm 0.004 \pm 0.009$    \\
      \hline 
    \end{tabular}
  \end{center}
  \label{tab:RelBF}
\end{table}

\section*{Acknowledgements}
%
%
\noindent We express our gratitude to our colleagues in the CERN
accelerator departments for the excellent performance of the LHC. We
thank the technical and administrative staff at the LHCb
institutes.
We acknowledge support from CERN and from the national agencies:
ARC (Australia);
CAPES, CNPq, FAPERJ and FINEP (Brazil); 
MOST and NSFC (China); 
CNRS/IN2P3 (France); 
BMBF, DFG and MPG (Germany); 
INFN (Italy); 
NWO (Netherlands); 
MNiSW and NCN (Poland); 
MCID/IFA (Romania); 
MICIU and AEI (Spain);
SNSF and SER (Switzerland); 
NASU (Ukraine); 
STFC (United Kingdom); 
DOE NP and NSF (USA).
We acknowledge the computing resources that are provided by ARDC (Australia), 
CBPF (Brazil),
CERN, 
IHEP and LZU (China),
IN2P3 (France), 
KIT and DESY (Germany), 
INFN (Italy), 
SURF (Netherlands),
Polish WLCG (Poland),
IFIN-HH (Romania), 
PIC (Spain), CSCS (Switzerland), 
and GridPP (United Kingdom).
We are indebted to the communities behind the multiple open-source
software packages on which we depend.
Individual groups or members have received support from
Key Research Program of Frontier Sciences of CAS, CAS PIFI, CAS CCEPP, 
Fundamental Research Funds for the Central Universities,  and Sci.\ \& Tech.\ Program of Guangzhou (China);
Minciencias (Colombia);
EPLANET, Marie Sk\l{}odowska-Curie Actions, ERC and NextGenerationEU (European Union);
A*MIDEX, ANR, IPhU and Labex P2IO, and R\'{e}gion Auvergne-Rh\^{o}ne-Alpes (France);
Alexander-von-Humboldt Foundation (Germany);
ICSC (Italy); 
Severo Ochoa and Mar\'ia de Maeztu Units of Excellence, GVA, XuntaGal, GENCAT, InTalent-Inditex and Prog.~Atracci\'on Talento CM (Spain);
SRC (Sweden);
the Leverhulme Trust, the Royal Society and UKRI (United Kingdom).

\clearpage

\addcontentsline{toc}{section}{References}
\bibliographystyle{LHCb}
\bibliography{main,standard,LHCb-PAPER,LHCb-CONF,LHCb-DP,LHCb-TDR}

\newpage
\centerline
{\large\bf LHCb collaboration}
\begin
{flushleft}
\small
R.~Aaij$^{38}$\lhcborcid{0000-0003-0533-1952},
A.S.W.~Abdelmotteleb$^{57}$\lhcborcid{0000-0001-7905-0542},
C.~Abellan~Beteta$^{51}$\lhcborcid{0009-0009-0869-6798},
F.~Abudin{\'e}n$^{57}$\lhcborcid{0000-0002-6737-3528},
T.~Ackernley$^{61}$\lhcborcid{0000-0002-5951-3498},
A. A. ~Adefisoye$^{69}$\lhcborcid{0000-0003-2448-1550},
B.~Adeva$^{47}$\lhcborcid{0000-0001-9756-3712},
M.~Adinolfi$^{55}$\lhcborcid{0000-0002-1326-1264},
P.~Adlarson$^{82}$\lhcborcid{0000-0001-6280-3851},
C.~Agapopoulou$^{14}$\lhcborcid{0000-0002-2368-0147}
% ...

{\footnotesize \it

% ...
$^{14}$Universit{\'e} Paris-Saclay, CNRS/IN2P3, IJCLab, Orsay, France\\
% ...
$^{38}$Nikhef National Institute for Subatomic Physics, Amsterdam, Netherlands\\
% ...
$^{47}$Instituto Galego de F{\'\i}sica de Altas Enerx{\'\i}as (IGFAE), Universidade de Santiago de Compostela, Santiago de Compostela, Spain\\
% ...
$^{51}$Physik-Institut, Universit{\"a}t Z{\"u}rich, Z{\"u}rich, Switzerland\\
% ...
$^{55}$H.H. Wills Physics Laboratory, University of Bristol, Bristol, United Kingdom\\
% ...
$^{57}$Department of Physics, University of Warwick, Coventry, United Kingdom\\
% ...
$^{60}$School of Physics and Astronomy, University of Glasgow, Glasgow, United Kingdom\\
$^{61}$Oliver Lodge Laboratory, University of Liverpool, Liverpool, United Kingdom\\
% ...
$^{69}$Syracuse University, Syracuse, NY, United States\\
% ...
$^{82}$Department of Physics and Astronomy, Uppsala University, Uppsala, Sweden, associated to $^{60}$\\
% ...
}
\end{flushleft}

\end{document}